\documentclass[aps,twocolumn,showpacs]{revtex4}
\usepackage{epsf}
\def\beq{\begin{equation}}
\def\eeq{\end{equation}}

\def\bk{{\bf k}}
\def\bk'{{\bf k}'}

\usepackage{graphics}
\usepackage{epsf}
 \usepackage{graphicx}
\def\be{\begin{eqnarray}}
\def\ee{\end{eqnarray}}

\def\bk{{\bf k}}
\def\bk'{{\bf k}'}

 \newcommand\beqn{\begin{eqnarray}}
 \newcommand\eeqn{\end{eqnarray}}

\begin{document}
%
\title{Hadronization and final state interaction effects in semi-exclusive Deep Inelastic Scattering off nuclei
}

\author{C. Ciofi degli Atti}
 \author{L.P. Kaptari}
 \altaffiliation{On leave from  Bogoliubov Lab. Theor. Phys.,141980, JINR,  Dubna, Russia}
  \affiliation{Department of Physics, University  of Perugia and INFN Sezione di Perugia,
       via A. Pascoli, Perugia, I-06123, Italy}
       \author{B. Z. Kopeliovich}
\affiliation{Max Planck Institut f\"ur Kernphysik Postfach 103980, 69029  Heidelberg, Germany}
%
%

\begin{abstract}
Recent calculations of the effects of  hadronization and  final 
state interaction (FSI) in semi-exclusive
 deep-inelastic scattering (DIS) $A(e,e'(A-1))X$ processes 
    are reviewed. The basic ingredient underlying these calculations, {\it viz}
   the time-dependent effective debris-nucleon cross section is illustrated, and some relevant results
   on complex nuclei and the deuteron are presented. In the latter case, particular 
 attention  is paid to the choice of the kinematics, for such a choice would in principle allow one to
 investigate both the structure function of a bound nucleon as well as the hadronization mechanisms.
 It is stressed that a planned experiment at Jlab on the process $D(e,e'p)X$ could be very useful in that respect. 
 \end{abstract}
%
\maketitle

\section{Introduction}
\label{intro}
To date, information on hadronization mechanisms  comes mainly from the
measurement of the multiplicity ratio of the lepto-produced hadrons in
semi inclusive $A(e,e'h)X$ processes \cite{HERMES},
whose interpretation on the basis of the quark re-interaction model of Ref. \cite{kope1} appears to be very convincing.
 However,
it should also be pointed out  that the more exclusive processes  of
 $A(e,e'(A-1))X$ (the {\it semi- exclusive}  process) proposed in Ref. \cite{tagging},
  though difficult to perform, could 
provide more direct information on quark re-interaction and
hadronization mechanisms, as shown recently in \cite{bocla}.
The semi-exclusive reaction $A(e,e'(A-1))X$  represents
the process in which an incoming electron   undergoes  a DIS
process off a low-momentum bound nucleon, with the scattered electron and the recoiling  $(A-1)$ nucleus
 detected in coincidence in the final state. Much theoretical attention has been devoted 
 to such process on a deuteron target, i.e.  when $A=D$ and 
 $(A-1)=p$ or $n$ (see e.g. Refs. \cite{simula}-\cite{alex}). These calculations
 (except Ref.\cite{wally}, to be discussed later on) are based upon the {\it plane wave impulse approximation
 (PWIA)}, according to which: i) $X$
results from DIS off  one of the two nucleons in the deuteron, ii) the second nucleon $N$
recoils without interacting with $X$ and is detected in coincidence with
the scattered electron. 
The process $A(e,e'(A-1))X$ on a complex nucleus was thoroughly investigated 
in  Ref. \cite{tagging} within the PWIA, i.e. by assuming
that the nucleon debris created by the
virtual photon  propagates   without re-interacting with the
spectator nucleus, which, therefore, always remains intact, an assumption
which, at first sight, might appear unjustified.

  As a matter of fact, as pointed out in Ref. \cite{bocla}, DIS  off a bound nucleon results in the
production of a multi-particle final state with an effective mass squared
equal to
 \begin{eqnarray}
s'&&\simeq m_N^2-Q^2 + 2\,m_N\,\nu -
2\,\sqrt{\nu^2-Q^2}\,p_L\nonumber\\
&&=Q^2(\frac{1}{x}-1)+m_N^2 -2|{\bf q}| p_L
\label{b.1}
 \end{eqnarray}
 where $Q^2={\bf q}^2 -{\nu}^2$ is the four- momentum transfer, $\nu$ the
virtual photon energy in the rest frame of the nucleus, $p_L$ the
longitudinal Fermi momentum of the nucleon relative to the direction of
the virtual photon (${\bf q}$ $\parallel z$), $m_N$ the nucleon mass, and
$x=\displaystyle\frac {Q^2}{2m_N\nu}$ the Bjorken scaling variable (we
neglect here the binding energy of the nucleon). At high energies and far
from the quasi-elastic region ($x\approx 1$), the effective mass is large,
$\sqrt{s'} \gg m_N$, and one  expects the  production of many particles,
which can interact traveling through the nucleus. This would
substantially suppress the probability for the spectator nucleus to remain
intact. However, the process of multi-particle production has a specific
space-time development, and it turns out that not so many particles have
a chance to be created inside the nucleus.
In order to check such an expectation,
in Ref. \cite{bocla}  
the propagation of the struck nucleon debris and its  re-interaction with the nuclear medium, 
 occurring  in the  semi-exclusive  process on complex nuclei 
 $A(e,e'(A-1))X$,   has been considered  with the aim of also clarifying if and 
  to which extent  the process is
sensitive to the details of quark hadronization in nuclear environment.
To this end   an effective time-dependent cross section $\sigma_{eff}$
(to be called the {\it debris-nucleon}
cross section) has been obtained, which 
describes the interaction with the nuclear medium of the hadronization products
 of a highly virtual quark created in a DIS process off a bound nucleon.  The main features of the 
  {\it debris-nucleon} cross section
are discussed in the next section.
 
 \section{The debris-nucleon cross section}
 In the derivation of the effective debris-nucleon cross section of Ref. \cite{bocla}
 use has been made of the color string model \cite{string} and  the gluon radiation
 mechanism \cite{gluon,kope1}, namely an approach has been adopted  which 
  takes into
account both the production of hadrons due to the breaking of the color
string  which is formed after a quark is knocked out off a bound nucleon,
as well as the production of hadrons originating from gluon radiation.

According to  QCD,  DIS is the process in which 
  an incident electron interacts with a  target quark by exchanging
  a gauge boson, making the quark  highly virtual.
   The formation of the final, detectable  hadrons, occurs after the  
   space-time propagation of the created nucleon debris,  
  with a sequence of soft and hard production  
  processes. The theoretical description of these  processes, which generally  cannot 
  be treated within
  perturbative QCD,  requires the use  of model approaches.
  Most of them are   based  upon  the quark color string model \cite{string}, 
  according to which
at  world interval of the order of $\simeq 1 fm$, \,  the string,  which is  formed by the   highly  virtual
  leading quark  and the remnant  target quarks,  breaks into  a hadron and another,
   less stretchy
  string. Further,  at longer space-time intervals,  this decay process iterates 
   unless the energy  of the string  is too low for   hadron 
   production and                       
  all the final hadrons are formed.  However, since the  hadronization process can
  also  
  be accompanied  by  gluon 
  perturbative bremsstrahlung \cite{gluon}, the string model itself is not sufficient for
  a consistent treatment of hadronization. A reliable 
  model must incorporate both the perturbative and the non-perturbative aspects of the 
   hadron formation process. 
   Note, that the hadronization process starts at extremely short space-time intervals, 
  hence a direct experimental study  of these intervals is difficult to undertake in DIS off a free nucleon.
  As a matter of fact,    
  the final hadrons do not carry much information
  about their early stage of hadronization, and therefore  only nuclear targets, which consist
  of a number of scattering centers, allow  one  to probe short times after
   the hadronization has started.
   The  time-dependent cross section $\sigma_{eff}$  describing  the scattering of the nucleon debris  
   with the surrounding medium turns out to read as follows \cite{bocla}

\be
\sigma_{eff}(t)=\sigma^{NN}_{tot} +
\sigma^{\pi N}_{tot}\Bigl[n_M(t) +
n_G(t)\Bigr]\ ,
\label{bb.6}
 \ee
  where $\sigma^{NN}_{tot}$ ($\sigma^{\pi N}_{tot}$) 
  are the total cross sections describing  nucleon-nucleon ( 
  meson-nucleon) interactions, and  $n_M(t)$ and $n_G(t)$ are the effective
  numbers of created mesons and radiated gluons, respectively, and 
  are explicitly given in Ref. \cite{bocla}. It should be pointed out that  in obtaining Eq. (\ref{bb.6})
   the color-dipole
   picture was employed by replacing each radiated gluon by a color-octet $q\bar q$
   pair.

\begin{figure}       
    \includegraphics[height=0.30\textheight]{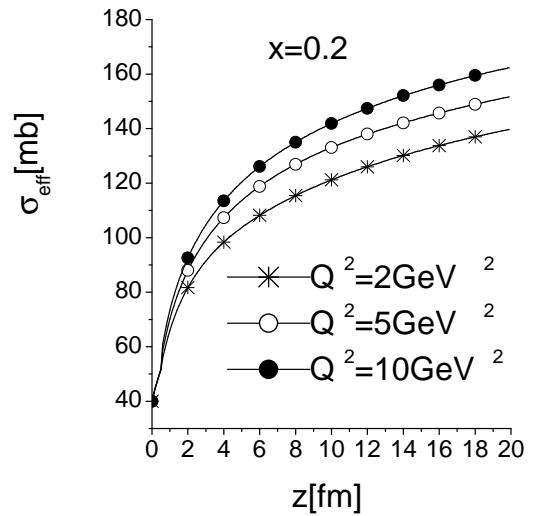}
    \vskip -0.1in
\caption{The debris-nucleon effective cross 
section (Eq. \ref{bb.6}) plotted {\it vs} the distance $z$ for a fixed 
value of the Bjorken scaling variable  
$x$ and various values of the four-momentum transfer 
 $Q^2$ (after  Ref. \cite{bocla}).}
 \label{Fig1}
 \vskip 0.1in
\end{figure}

 The cross section (\ref{bb.6}) exhibits a rather complex 
 $Q^2$- and $x$-dependence, which, however,   asymptotically  
 tends to a  simple logarithmical behavior.  This is illustrated  
 in Fig.~\ref{Fig1}, where
 the dependence of $\sigma_{eff}$  upon the coordinate $z$ along the propagation direction is shown 
for different values of the 
  four momentum transfer  $Q^2$ and a fixed value of the Bjorken scaling variable $x$.
 
 It should be stressed that in Ref. \cite{bocla}  it has been shown  that the effective cross section
 (\ref{bb.6})
 can reasonably explain the Fermilab data on the production
 of protons in semi inclusive Deep Inelastic muon-Xenon scattering 
 at $490\,\, GeV$  \cite{adams}.  In this experiment, at a given value of $x$,  
 {\it grey tracks} were observed, which have been 
  interpreted as protons in the momentum  range
 $200<p<600\,\, MeV/c$. Using  (\ref{bb.6})
 the average number $<n_g>$ of {\it grey tracks} {\it vs}\,\, $Q^2$ has been calculated, obtaining a
  satisfactory explanation
 of the experimental data of Ref. \cite{adams}; this  makes us  confident in the correctness of the
 calculation of FSI effects in semi-exclusive processes off nuclear targets, to be discussed in the next Section.

 \section{The process $A(e,e'(A-1))X$ in complex nuclei}  
 
 In a nucleus, at each hadronization point one  expects 
  re-interactions of the produced hadrons  with the nuclear constituents, 
  so that  the multiplicity of final particles is predicted to be reduced relative
  to the case of nucleon targets.
  \begin{figure}[h] 
\includegraphics[height=0.40\textheight]{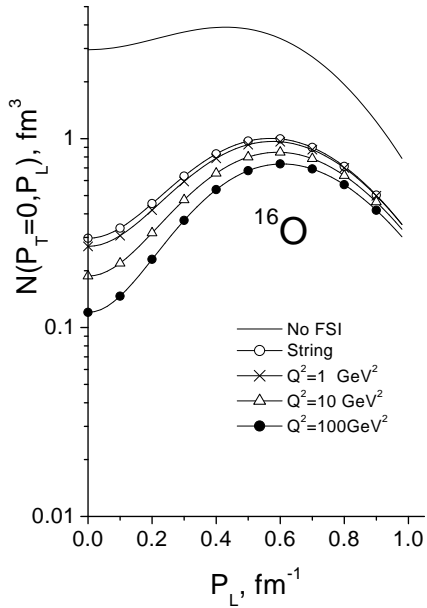}
{\caption[Delta]
 {The proton Momentum Distribution for $^{16}O$  (Eq. (\ref{mom}) - No FSI)
compared with the Distorted Momentum Distribution $N({\vec P}_{A-1}$)  
(Eq.(\ref{nondi})) plotted {\it vs} $P_L$ for $P_T=0$. The
curve labeled by open dots has been obtained using the effective
 cross section for the nucleon debris corresponding to the color string model,
  whereas the other curves correspond to the cross
section  which includes also the gluon
bremsstrahlung. The stars represent the distorted {\it proton} momentum
distributions calculated in Ref. \cite{hiko2} for the semi-inclusive quasi-elastic process
$^{16}O(e,e'p)X$  (After Ref. \cite{bocla}).}
\label{Fig16O}}
 \end{figure}
  Thus, by comparing the same DIS process off a single nucleon
  and off nuclear targets, information on the space-time structure of the hadronization 
  process could be obtained.
    As already pointed out, the  theoretical model of  hadronization developed in Ref.\cite{kope1},
  proved to be very effective for the explanation  of the  leading hadron  multiplicity ratios (nucleus
  to nucleon)  measured at HERMES
   \cite{HERMES}  in semi-inclusive processes.
  It should however  be pointed out that   the initial stage of hadronization
  is difficult to investigate  by semi-inclusive  processes, where  the non leading
  hadrons are strongly affected by subsequent cascade processes and therefore do not carry 
  information  on  their  formation mechanism.
 
\begin{figure}          
\epsfxsize 3in
\epsfbox{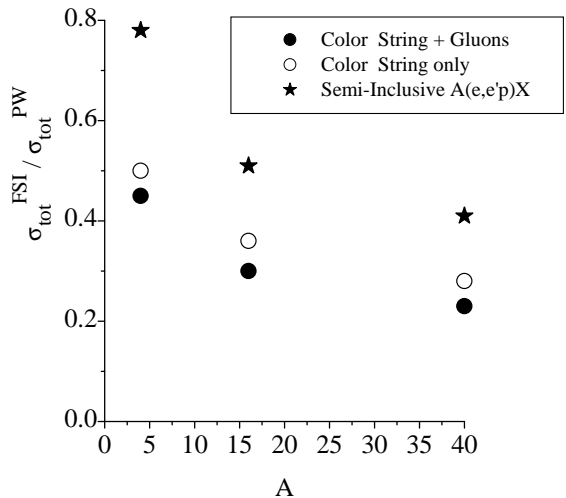}
 
\caption{ {The nuclear transparency i.e. the ratio between the total cross section $\sigma_{tot}^{FSI}$,
obtained by integrating Eq.~(\ref{nondi})   over ${\vec P}_{A-1}$) and the PWIA total cross section
$\sigma_{tot}^{PW}$, 
obtained  in the same way but 
  disregarding 
  the Final State Interaction ($S(\vec b, z) = 0$, $\sigma_{tot}\equiv \sigma_{tot}^{PW}$). 
  The open
dots correspond to the debris-nucleon effective cross section  given by
the color string model, whereas the full dots
correspond to the cross section where the gluon bremsstrahlung has also
been considered [Eq.~(\ref{bb.6})] at $Q^2=100GeV^2$. The stars represent
the {\it proton} transparency calculated in Refs.~\cite{hiko1,hiko2} for the
reaction $A(e,e'p)X$ } (After Ref. \cite{bocla}).}
 \label{Fig8}
\end{figure}

   In Ref. \cite{bocla} it has been shown 
   that
  the deep inelastic semi-exclusive process $A(e,e',(A-1))X$, where the nucleus $(A-1)$ is detected  
  in coincidence with the scattered electron,  could  be  an effective tool to study
    the mechanisms and the initial stage of  hadronization.
    The approach of ref. \cite{bocla} is based on a traditional Glauber-type nuclear structure approach in which
the ground state wave function of the initial nucleus 
$\Psi_A^0 ({\vec r},{\vec r}_2\dots{\vec r}_A)$ is written as a product
of a function $\phi$, describing the motion of the hit nucleon and the wave
function $\Psi_{A-1}^f({\vec r}_2\dots{\vec r}_A)$ of the spectator; moreover  
$|\Psi_{A-1}^f({\vec r}_2\dots{\vec r}_A)|^2$ is, as usually,  approximated by  a
product of single particle densities. Within these assumptions, the cross section can be shown to 
be governed by the {\it distorted momentum distributions}
 \beq
n_A^{FSI}(\vec P_{A-1})\equiv N(\vec P_{A-1})=\left| F_{A,A-1}^{FSI}({\vec
P_{A-1}})\right |^2\ 
\label{nondi}
 \eeq
where
 \beqn
F_{A,A-1}^{FSI}
\simeq \int
e^{i{\vec P}_{A-1}\,\vec r}\,
\phi ({\vec r})
\left[1 - \frac{S({\vec b},z)}{2(A-1)}\right]^{A-1}\, 
d {\vec r}\ 
\label{b.8b}
 \eeqn
and
 \beq
S({\vec b},z)=\int\limits_{z}^{\infty} dz'\,
\rho_A({\vec b},z')\,\sigma_{eff}(z'-z)\ 
\label{b.9}
 \eeq
 \noindent In these equations $\rho_A({\vec b},z)$ is the nuclear density ($\int d{\vec r}
\rho_A(\vec r)=A$), and
$\sigma_{eff}(z'-z)$ is given by  (\ref{bb.6}). Note that when FSI are absent, i.e. when
$\sigma_{eff} = 0$, the usual nucleon momentum distribution in the nucleus is recovered
\beq
n_A(|\vec P_{A-1}|)\simeq
\left| \int
e^{i{\vec P}_{A-1}\,\vec r}\,
\phi ({\vec r})\, 
d {\vec r}\right |^2 
\label{mom}
 \eeq
The distorted momentum distributions for $^{16}O$
are shown in Fig. \ref{Fig16O}, whereas the nuclear transparency for  the nucleon debris created in a DIS process
 and for a nucleon knocked out in
 a quasi-elastic scattering semi-inclusive $A(e,e'p)X$ process, is exhibited in Fig. \ref{Fig8}.

    \section{The process $D(e,e'p)X$}
To perform an experiments on semi-exclusive processes  $A(e,e'(A-1))X$ off complex nuclei is not easy task;
however, in the case of  a deuteron target the experiment appears to be feasible \cite{kuhn}. As a matter of fact
the  process
 \begin{equation}
e+ D = e'+ X + N
\label{reaction}
\end{equation} 
 has been the object
of many theoretical calculations, mainly aimed at studying the neutron structure function
\cite{simula}-\cite{alex}, and  its experimental 
investigation is planned to be performed at JLab \cite{kuhn}.
  
  Process (\ref{reaction}) has many attractive features with respect to the inclusive process
  $^2H(e,e')X$, both for extracting information on the  nucleon structure functions and for the
  investigation of  hadronization mechanism.
  Indeed,
  in spite of the fact that
  inclusive DIS processes   have  provided us in the past  with 
  fairly precise knowledge of parton distributions in hadrons, conclusive
  information about the origin of the EMC effect is still lacking; moreover, important details
  on the neutron structure function are unknown, which  is 
  mostly due  the difficulties and ambiguities  related to the  unfolding of the neutron
  structure functions from  nuclear data \cite{ambig}. Semi-exclusive 
  processes could provide, on the contrary, unique information on both the origin
  of the EMC effect, and  the details of the neutron structure function;    
  moreover, they  can also be used as  a  unique tool to investigate   
  hadronization processes. Obviously, a reliable treatment of semi-exclusive processes
  requires a careful treatment of  the  FSI of the nucleon  debris $X$ with the final nuclear system $(A-1)$.
  Intuitively, one expects, on one hand,  that   if the proton is detected
in the backward hemisphere, FSI effects should not play a relevant role, so that
the process could be used to investigate the structure functions of bound nucleons;  on  the other hand,
the effects from FSI are expected to be relevant in the process when the
 recoiling nucleon is detected  in the direction perpendicular to the three-momentum
transfer, in which case information on the  hadronization mechanism could be obtained.
In view of the planned experiments at JLab \cite{kuhn}, a detailed quantitative calculation of FSI effects in process
(\ref{reaction}) has been carried out in Ref. \cite{CKK}. There it
 has been found that the central quantity describing the process is the
distorted momentum distribution
 \beqn
 &&n_D^{FSI}({\bf p}_s,{\bf q}) =\\\nonumber
&&\frac13\frac{1}{(2\pi)^3} \sum\limits_{{\cal
M}_D} \left | \int\, d  {\bf r} \Psi_{{1,\cal
M}_D}( {\bf r}) S( {\bf r},{\bf q}) \chi_f^+\,\exp (-i
{\bf p}_s {\bf r}) \right |^2
 \label{dismomfsi}
\eeqn
where $\chi_f$ is the  spin function of the spectator nucleon and 
$S( {\bf r},{\bf q})$   the $S$-matrix describing 
the  final state interaction
between the debris and the  spectator,  {\it viz.}  
\beqn
&&S({\bf r},{\bf q}) =\\\nonumber
&& 1-\theta(z)\, \frac{\sigma_{eff}(z,Q^2,x)(1-i\alpha)}{4\pi b_0^2}\,
\exp(-b^2/2b_0^2).
 \label{gama}
\eeqn

When FSI are absent ($\sigma_{eff}=0$) the usual deuteron momentum distributions are recovered, {\it viz}
\beqn
 &&n_D(|{\bf p}_s|,) =\\\nonumber
&&\frac13\frac{1}{(2\pi)^3} \sum\limits_{{\cal
M}_D} \left | \int\, d  {\bf r} \Psi_{{1,\cal
M}_D}( {\bf r})\chi_f^+\,\exp (-i
{\bf p}_s {\bf r}) \right |^2
 \label{momD}
\eeqn

\begin{figure}
  \begin{center}          
\epsfxsize 3.0in
\epsfbox{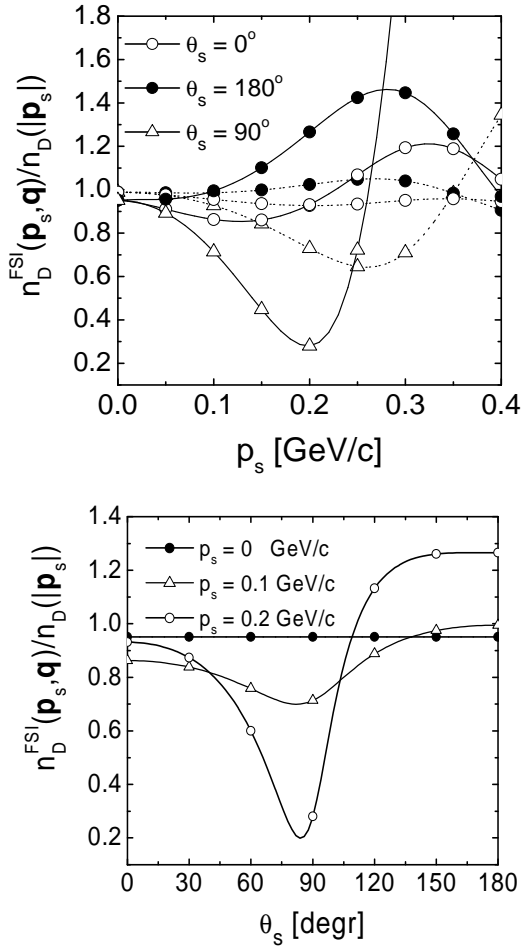}
\end{center}
 \caption{ {\it Upper panel}: the Deep Inelastic Scattering ratio  $n_D^{FSI}/n_D$
for the process $D(e,e'p)X$ with $n_D^{FSI}$ and $n_D$  (given, respectively, by Eqs. (8) and (10)),
 calculated
{\it vs.} the momentum  $p_s \equiv |{\bf p}_s|$ of the spectator nucleon emitted at different angles $\theta_s$.
The full lines correspond to the  $Q^2$- and $z$-dependent   debris-nucleon
effective cross section $\sigma_{eff}$ shown in Fig. \ref{Fig1}, whereas the
 dashed lines correspond to a constant cross section $\sigma_{eff} = 20\,\, mb$. {\it Lower panel}: the same as in 
 the upper panel, {\it vs} the  emission angle ${\theta_s}$ of the spectator, for different values of the
 spectator momentum. Calculations have been performed at  
 $Q^2=5\,\, (GeV/c)^2$ and $x=0.2$ (After Ref.\cite{CKK}).}
 \label{Fig2}
\end{figure}

In what follows, the  momentum distributions  are  expressed in terms of  
 the light cone variable $\alpha_s$, $p_{\parallel}$ and $p_{T}$, defined as
\be
\alpha_s=\frac{E_s-p_{\parallel}}{m},\quad
p_{\parallel}= |{\bf p}_s|\cos \theta_s ,\quad
    p_{T}= |{\bf p}_s|\sin \theta_s 
\label{eq8}
\ee
\noindent where $\theta_s$ is the angle between ${\bf p} _s$ and ${\bf q}$, 
 and the spectator four-momentum is $p_s\equiv (E_s, {\bf p}_s)$ 
 (note, that in the  DIS kinematics the  light cone $z$-axis is directed opposite to the
vector ${\bf q}$). 
In Ref. \cite{CKK} the problem  was addressed of finding  proper kinematics which would allow the
investigation of both the nucleon structure function and the hadronization mechanism; in the first case 
 FSI effects should be minimized, whereas in the second case they have to be maximized.
 Since   all of the FSI effects are contained  in the
 distorted momentum distribution  $ n_D^{FSI}( {\bf p}_s, {\bf q})$,
  its deviation from the 
 nucleon momentum distributions  $n_D (|{\bf p}_s|)$ could 
   provides clear 
   signature of  FSI effects. This is   illustrated in Fig. \ref{Fig2}, which shows
 the ratio
$n_D^{FSI}/n_D$,  calculated   using two  different 
models for the effective cross section $\sigma_{eff}$, representing its  
upper and lower limits, {\it viz} the time- and $Q^2$-
dependent cross section  of Ref. \cite{bocla} (solid lines), and 
a constant cross section $\sigma_{eff}=20\,\, mb$ 
(dashed lines) considered in Ref. \cite{wally}. 
In our  numerical  calculations the parameters entering  Eq. (\ref{gama}), {\it viz}
the slope  $b_0$  and the ratio $\alpha$ of the real to the
imaginary parts of the forward amplitude, have been  taken from $NN$ 
scattering data
at high energies.
It should be pointed out, in this respect, that our results, in the range of considered momenta, 
are not very sensitive to the value of $\alpha$; whereas the values of the latter 
is known in the case of {\it nucleon-nucleon} scattering, the value for  
{\it debris-nucleon} scattering  should rely on some theoretical models. This
point is under investigations and the results will be presented elsewhere.
It can be seen from 
Fig.~\ref{Fig2}
 that the predictions given by the two different 
models of the effective cross section ere rather  different,
 particularly when the recoiling proton is emitted perpendicularly to ${\bf q}$, i.e. 
 at $\theta_s \sim 90^o$,  and with large values of the 
momentum ($p_s\sim 0.2\,\, GeV/c$). 
 Thus, by investigating this  kinematical  region  one could, in principle,  obtain 
unique information about the magnitude of  $\sigma_{eff}$
and, consequently, about the hadronization mechanism.
 
\begin{figure}[t] 
\includegraphics[height=0.6\textheight]{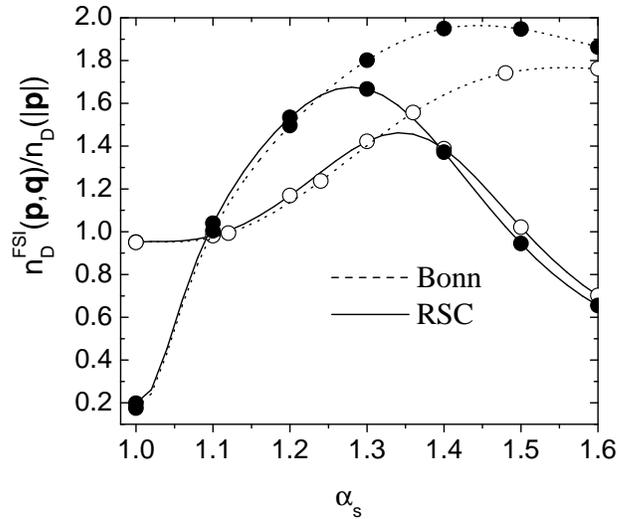} 
\vskip -6cm
\caption{The Deep Inelastic Scattering  ratio $n_D^{FSI}/n_D$ for the process $D(e,e'p)X$ {\it vs} the light cone variable 
 $\alpha_s$,
 calculated at $Q^2=5\,\, GeV^2/c^2$
and $x=0.2$ using two different deuteron wave functions: the solid lines correspond to 
the  RSC potential and the dashed lines  to the  Bonn  
potential. Calculations have performed in correspondence of two values of the transverse
momentum $p_T$ of the recoiling nucleon  (cf. Eq. \ref{eq8}), namely  $p_T= 0 \,  GeV/c$ (open dots) and
$p_T= 0.2\  GeV/c$ (full dots). The results correspond to the debris-nucleon cross section  shown in Fig.1
 (After Ref. \cite{CKK}).}
\label{Fig4}  
\end{figure}

The results exhibited in Figs.  \ref{Fig2}  also demonstrate that FSI effects are essentially reduced
in  parallel kinematics ($\theta=0^o, 180^o$)  and also at small values of $p_s$. 
It can be shown that in this region the cross section provides direct
information on the  structure function $F_2$ of a bound nucleon, so that a reliable investigation
of  off-mass  
shell effects \cite{kaptoff,thomas} in DIS could be possible. The results exhibited in Figs. \ref{Fig2} 
 have been
obtained by using the deuteron wave function corresponding to  the Reid Soft Core (RSC)
 potential \cite{RSC}. Calculations have been repeated with the Bonn interaction \cite{bonn}.
As shown in Fig. \ref{Fig4}, at small values of $\alpha_s$ different  potential models 
yield  basically the same  results; however at  moderate and large values of $\alpha_s$ the predictions by
different deuteron wave functions appreciably differ.

To sum up, from the analysis  exhibited in Ref. \cite{CKK}, it can be  concluded  that FSI effects in
   the semi-exclusive process
  (\ref{reaction}) are negligible  in the backward kinematics and slow
   momenta of the detected nucleon, which would allow one to investigate  the nucleon 
   structure function of bound nucleons, in particular the neutron one;  if,  on the contrary,
   the spectator nucleon is emitted perpendicularly to the momentum transfer, FSI effects
   are enhanced and different models for the hadronization process could be investigated.
   
   \section{Summary and Conclusions}
   We have reviewed recent calculations of FSI effects which occur in the semi-exclusive 
    deep inelastic $A(e,e'(A-1))X$ process. The FSI
   is generated by the rescattering with the medium of the nucleon debris formed by the deep inelastic scattering of the
   virtual photon. These calculations are based upon the effective time-dependent
   debris-nucleon cross section obtained in Ref. \cite{bocla} on the basis of the color string \cite{string} 
   and the gluon radiation \cite{gluon} models. The detailed calculation \cite{CKK} of the reaction
   $D(e,e'p)X$, shows that the planned experiment at Jlab \cite{kuhn}, aimed at investigating such a reaction,
   may provide unique information on the hadronization mechanisms.

   \section{Acknowledgments}
   This work was partially supported by the Ministero della Istruzione, Universit\`{a} e Ricerca (MIUR), 
 through the funds COFIN01.
L.P.K. is indebted to  the University of
Perugia and INFN, Sezione di Perugia, for warm hospitality and financial support.


 \end{document}